# Research Director's Report

Sakue Yamada

KEK – ILC Promotion Office
Tsukuba, Ibaragi – Japan

Recent development of the detector activities and related events are reported and some details of the newly completed management structure are described. Also the plans for processes after the LOI due date are introduced. [1]

## 1  Project Advisory Committee

The Project Advisory Committee (PAC) was established by ILCSC newly. It oversees the organization and activities of both the GDE and the detector management. The first PAC meeting was held on October 19 and 20 in Paris at LPNHE, University of Pierre et Marie Curie. This time, we wished mainly to have our organization and direction of activity examined so that we can be sure to be on right tracks. Presentations were made on the general detector activity and organizational issues by me, on IDAG activities by its chair Michel Davier, on the MDI matters by Andrei Seryi of GDE, and on ILC-CLIC cooperation by Francois Richard. At the end of the meeting PAC commended the present scheme of the management, IDAG's composition and its start of activity and also the cooperation with the CLIC detector group. PAC recommended ILCSC to extend the mandate of IDAG through the Technical Design Phase to monitor LOI groups' development. It also recommended that the detector community makes a written interim progress report at the end of TDP-I in 2010.

## 2  ILCSC

At the ILCSC meeting held at SLAC on October 31, IDAG mandate was extended as PAC recommended till 2012 through to the end of Technical Design Phase. The entire schedule and plan of the LOI process was modified by ILCSC, February last year, and this redefinition of the mandate for IDAG which is in accordance with the new plan clarified remaining uncertainty.
 I reported the nearly completed management structure of the detector activity with most names filled in. This list became complete after the ILCSC meeting as shown at the workshop and described below.
 There was a presentation by Dr. H. Sugawara to propose inclusion of a photon-photon collider, as the first phase option, into the presently going design effort. After some discussions, ILCSC wished GDE and the Physics Panel of the detector management to prepare more material for their next meeting so that they can consider more in details. Following this, GDE formed a small group to look into accelerator related issues and I asked the convener of the Physics Panel to investigate the physics case. (During this workshop these subgroups met together and decided to collaborate to produce a unified document.)

## 3 Management Organization

Regarding the management organization, we reconsidered the present membership of the Regional Contacts who are WWS co-chairs. When the management was formed in 2007, it was extremely helpful to keep close contact with the detector community in order to start the activity. After one year from the start, we made a good progress and the management structure was completed. However, I believe the intimate link with the community is still important particularly when the environment around us is not very stable. This stance was commended by PAC and was supported by ILCSC as well as by WWS OC.

Shortly before the LCWS08 meeting we could fill the names of the entire members and conveners of the common task groups to complete the management structure. The common task groups will conduct many works which are common to all the LOI groups and form key elements of the detector activity. There are five common task groups; Machine Detector Interface (MDI) group, Engineering Tools group, Detector R&D Panel, Software Panel and Physics Panel. Their members can be found in the ILC web page under Detector and Experiments. In general the common task groups are made of members from the three LOI groups. However, for the Detector R&D Panel and the Physics Panel we invited more members widely from related physics communities. I will describe each group in the following.

With the conveners of the common task groups, the Physics and Experiment Board of the detector activity management became complete. It consists of the Executive Board members, representatives of the LOI groups and the conveners of the common task groups, During this workshop the first meeting of the Physics and Experiment Board was made in face to face. Also each common task group met face-to-face during the workshop. Except for the MDI group which was active since April 2008, these were the first meetings and were useful to identify the mandate of the groups among the members and to discuss plans how to start.

### 3.1 Machine Detector Interface group

This group cooperates with the beam delivery system (BDS) group of the GDE to communicate on the parameters and matters regarding the interfacing area. It will cover a wide range of topics including the beam optics and experimental hall structure. The group is convened by Karsten Buesser and the deputy is Phil Burrows. Now they are working on a document for common understanding of minimum requirement.

### 3.2 Engineering Tools group

This group will communicate with the relevant members of the GDE in order to use common or compatible tools of engineering which will become necessary in the future to exchange design drawings. The convener of the group is Catherine Clerc. This group will also contact closely with the MDI and BDS groups, too.

**3.3 Detector R&D Panel**

This group was designed to be the largest in the number of members and was reinforced further by inviting members from detector R&D collaborations. Considering that the role of the R&D collaborations is very important when the LOI groups try to conduct their detector R&D for their concepts, it is extremely helpful and necessary to keep intimate cooperation with them. In that sense we are very glad that many R&D collaborations were willing to send their representatives to this common task group. Having the additional members will enable the group to stretch its view and activity beyond the scope of the LOI process. The convener is Marcel Demarteau and the deputy is Franco Grancagnolo.

**3.4 Software Panel**

This group will work on several software related issues like simulation codes, data sets and so on to be used commonly by the LOI groups. At present, some members of the group are participating intensively in the cooperative discussions with the CLIC detector group. The convener is Akiya Miyamoto and the deputy is Norman Graf.

**3.5 Physics Panel**

The group will consider physics related issues. Therefore it is natural that the group is enforced by inviting members from wider communities. Based on the recommendation of regional ILC communities, we invited one theorist and experimentalist from each region. The convener of the group is Michael Peskin and the deputies are Keisuke Fujii and Georg Weiglein. This group will first work on possible ILC physics and may develop its activity beyond that.

In principle, each group is free to organize its plan as long as it is useful for the ILC detector activity. Depending on the group, the scope may expand over the ILC time scale. When its focus remains on the ILC activity for the moment, we welcome such wider views. WWS used to have several panels of which roles were the same or close to those of the common task groups. Considering this overlapping, WWS decided to stop their panels during the last ECFA -LC workshop in Warsaw in the view that the same function would be carried by the new common task groups. It does not mean straightforwardly that the same functions, which sometimes were wider, were inherited. It is, however, not excluded that the roles of the common task groups are enlarged as time goes.

# 4 Cooperation with CLIC

Since about a year, cooperation with CLIC detector group is included in the general ILC-CLIC cooperation program between the GDE and CERN. The detector activity is listed as one of the five working groups. Regarding detector design or physics simulation, ILC and CLIC communities are in different stages at present. ILC community has developed several detector concepts and accumulated a large amount of simulation tools, data sets and experiences based on them. They are very useful for the CLIC community which is about to start similar activity. On the other hand CERN has rich experience of building extremely complex and large detector systems for LHC. The knowhow of constructing these detectors will be valuable to the ILC detector community. Very fruitful cooperation has started which are appreciated by the both sides.

## 5   Schedule from now

   Now every concept group is working hard to prepare for its Letter of Intent, of which due date is end March 2009. IDAG will be prepared to start validation process as soon as Letters of Intent are submitted.[2] The concept groups will present their LOIs during the next workshop in Tsukuba (TILC09, April 17-21) where IDAG will also meet. IDAG plans to come to a conclusion in autumn, 2009. There will be an additional meeting of IDAG during the summer apart from any ILC related workshops. This will be for detailed discussions between IDAG and individual LOI group. The validation includes the examination of the physics aims, adequacy of the chosen detector system, its performance e.g. regarding the benchmark reactions, feasibility including each component and the push-pull mechanism as well as capability of the group to conduct the foreseen design works through the technical design phases. IDAG issued some additional questions on top of those given in the LOI guideline document. They include questions on e.g. push-pull mechanism or accelerator background. More information about IDAG validation was explained by the IDAG chair at this workshop. Details can be found in his presentation.
   One important aspect of the LOI is that the groups are supposed to set priorities for their detector R&D program based on their concepts. This will be important for requesting budget for the R&D programs. Such planning may require detailed discussions with the detector R&D collaborations. We hope the groups can go through this procedure by the time of LOI submission and conduct necessary R&Ds through the Technical Design Phase-I.

## 6   References

[1] Slides of RD's Report
http://ilcagenda.linearcollider.org/getFile.py/access?contribId=5&sessionId=0&resId=0&materialId=slides&confId=2628

[2] Slides of IDAG chair
http://ilcagenda.linearcollider.org/getFile.py/access?contribId=55&sessionId=2&resId=0&materialId=slides&confId=2628